\documentclass[aps,prl,twocolumn,groupedaddress,superscriptaddress,colorlinks,bookmarks=false,citecolor=darkspringgreen,linkcolor=darkolivegreen,urlcolor=navyblue]{revtex4-1}

\usepackage{graphicx}
\usepackage{amsfonts}
\usepackage{amssymb}
\usepackage{amsmath}
\usepackage{dcolumn}
\usepackage[normalem]{ulem}
\usepackage{enumerate}
\usepackage{bbold}
\usepackage{cancel}
\usepackage{mathtools}
\usepackage{siunitx}
\usepackage{braket}
\usepackage{amsmath}
\usepackage{mdframed}
\usepackage{dsfont}
\usepackage[caption=false]{subfig}
\usepackage{soul}
\usepackage{orcidlink}
\usepackage{physics}
\usepackage{stackrel}
\usepackage{algorithm}
\usepackage{algpseudocode}

\definecolor{darkgreen}{RGB}{0, 159, 117}
\definecolor{navyblue}{rgb}{0.0, 0.0, 0.5}
\definecolor{darkolivegreen}{rgb}{0.33, 0.42, 0.18}
\definecolor{darktangerine}{rgb}{1.0, 0.66, 0.07}
\definecolor{darkspringgreen}{rgb}{0.09, 0.45, 0.27}
\definecolor{bordeaux}{rgb}{0.53, 0, 0.51}
\definecolor{reviewblue}{rgb}{0.25, 0.50, 0.80}

\usepackage{ mathrsfs }

\newcommand{\threestate}[3]{%
\begin{matrix}
#1\\[-.25cm]
#2#3
\end{matrix}
}

\newcommand{\twostate}[2]{%
\begin{matrix}
#1\\[-.25cm]
#2
\end{matrix}
}

\newcommand{\kett}[1]{\lvert #1 \rangle}
\newcommand{\brat}[1]{\langle #1 \rvert}

\usepackage{tikz}

\newcommand{\halfcirc}{%
  \mathbin{%
    \tikz[baseline=-0.5ex]{
      \draw[line width=0.07ex] (0,0) circle (0.42ex);
      \fill (0,-0.42ex) arc (-90:90:0.42ex) -- cycle;
    }%
  }%
}

\newcommand{\papertitle}{
Nonequilibrium phases and quantum correlations in 
synthetic transport models
}

\begin{document}

\title{\papertitle}

\author{Uddhav Sen}
\affiliation{
Centre for Fluid and Complex Systems, Coventry University, Coventry, CV1 2TT, United Kingdom}

\author{Federico Carollo}
\affiliation{Dipartimento di Fisica, Sapienza Università di Roma, Piazzale Aldo Moro 5, 00185 Rome, Italy}
\affiliation{
Centre for Fluid and Complex Systems, Coventry University, Coventry, CV1 2TT, United Kingdom}

\author{Sascha Wald}
\email{sascha.wald@coventry.ac.uk}
\affiliation{
Centre for Fluid and Complex Systems, Coventry University, Coventry, CV1 2TT, United Kingdom}

\begin{abstract}
    Quantum devices featuring mid-circuit measurement and reset capabilities, such as quantum computers and dual-species Rydberg quantum simulators, enable the realization of {\it quantum cellular automata}.
    These systems evolve in discrete time following local updates implemented by unitary gates, and allow for the realization of both closed and synthetic dissipative dynamics.
    Here, we focus on quantum cellular automata that implement minimal models of classical and quantum transport.
    To illustrate our ideas, we focus on a discrete-time totally asymmetric simple exclusion process and investigate how coherent dynamical contributions allow for the emergence of quantum effects and correlations.
    We find that bipartite entanglement dominates the transient evolution, while stationary states can retain quantum correlations beyond entanglement.
    Our results suggest viable routes for realizing transport models on quantum devices and characterizing collective quantum correlations in strongly driven systems.
\end{abstract}

\maketitle

{\bf Introduction.---}Transport phenomena provide a unifying framework for the study of nonequilibrium systems. 
At the microscopic scale, particle transport is often modeled using {\it simple exclusion processes}, which describe the stochastic dynamics of mobile agents on lattices and complex networks.
These processes have found a wide range of applications including RNA polymerization~\cite{MacDonald68,Lakatos2003,Chowdhury08,Scott2019,Juraj23}, transport through networks~\cite{Neri11,Bun14}, and traffic modeling~\cite{Nagel96,Bonnin2022}. 
Among the many variants, the asymmetric simple exclusion process (ASEP) and its totally asymmetric counterpart (TASEP) have emerged as  paradigmatic nonequilibrium models, characterized by a rich phenomenology, multiple stationary phases, with the stationary state exactly solvable in the classical case~\cite{Prol16,Ada07,Gor12,Gier11,Dep04,Brank00,Bun24,Blythe07,Derrida_1993,Derrida98,Schtz1997,Golinelli2006,Rkos2005,Spohn2006,Raj97,Chou2011,Hel01,Kavanagh_2022,Temme_2012,Blythe07}.

\begin{figure}[t]
    \centering
    \includegraphics[width=\columnwidth]{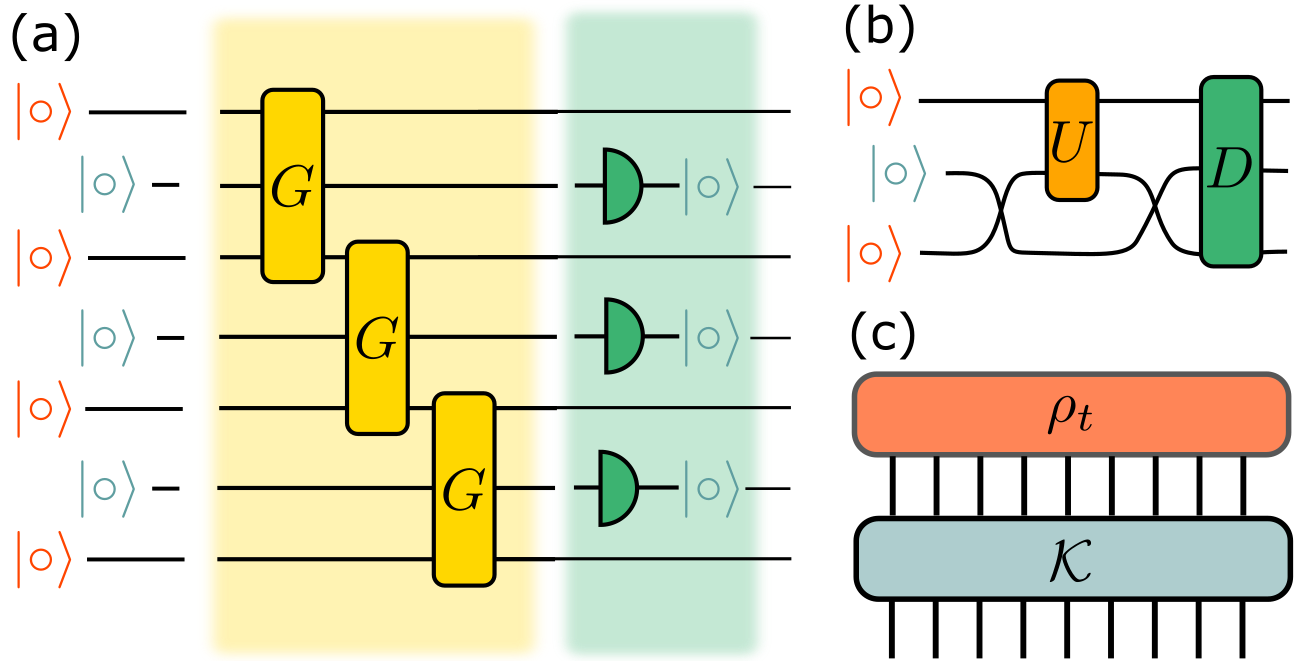}
    \caption{ {\bf Quantum cellular automaton from dual-species atom arrays.}  (a) Two-species atom array (sketched on the left) modeled of system qubits (left most in red) and ancillary qubits (right most in green).
    System and ancillary qubits interact through the repeated application of the local quantum gate $G$. 
    After a full sweep throughout the lattice, the ancilla qubits are measured and reinitialized and the sweep is repeated.
    (b) The gate $G$ is composed of two unitary gates: $U$ encodes coherent interactions within the system, $D$ induces ancilla-mitigated dissipation in the system. 
    SWAP operations are inserted to stress the nearest-neighbor range of the gate $U$ and to ensure the final qubit order is consistent with the initial qubit order; they do not affect the effective transport dynamics.
    (c) The full QCA update, consisting of sweep and ancilla reinitialization, implements a dissipative dynamics that can be expressed as a global Kraus map ${\cal K}$ that propagates the quantum state through time $\rho_{t+1} = {\cal K} [\rho_t]$.
    }
    \label{fig:sketch}
\end{figure}

{\it Cellular automata} (CA) are minimal discrete-time models whose fundamental units, so called cells, are updated according to local rules depending on the state of the neighboring cells~\cite{Wolf83,Mart84}.
Famous examples of CA include Conway's Game of Life~\cite{Gard70} or Wolfram's Rule 30 automaton~\cite{Wolf83,Wolf85}, both of which are of particular interest as they produce complex behavior from seemingly simplistic rules.
Their ability to generate emergent complexity motivates the study of CA dynamics and makes them a versatile tool for designing synthetic dynamical systems.
Due to their locality, CA provide a natural minimal framework for modeling simple exclusion processes.

CA have also been extended to the quantum setting.
Here, each cell consists of a qubit (or more generally a $d$-level quantum system, known as a qudit), and the update rules are replaced by local unitary gates~\cite{Farrelly2020}, see Fig.~\ref{fig:sketch}(a).
Quantum CA (QCA) have recently been realized experimentally in programmable atomic arrays~\cite{White2026}.
Mid-circuit measurements and qubit resets enable the implementation of incoherent stochastic processes, allowing to engineer synthetic open-system dynamics. 
In these settings, QCA have been identified as a potential approach to overcoming challenges associated with reliable manipulation of large many-qubit systems as their simple structure can limit the complexity of the coherent dynamical control~\cite{Cesa2026,White2026}.
Furthermore, QCA have been utilized, both theoretically and experimentally, to explore nonequilibrium dynamics~\cite{Duranthon21,Gill20,sfairopoulos25} as well as phase transitions into absorbing states~\cite{Lesa19} and collective properties in quantum neural networks~\cite{Wintermantel20,Gill23,Boneberg_24,Boneberg_25}.
However, QCA have not been used for exploring quantum transport, which is the main purpose of this paper, particularly since discrete-time dynamics can lead to qualitatively different correlation structures~\cite{Wald2025}.

Recently, quantum versions of simple exclusion processes have attracted significant interest~\cite{Albert2026, Bernard21, Bernard22, Bernard25, Alba2025,Russotto2025}. 
Notably, the TASEP has emerged as a fundamental building block of {\it active quantum matter}~\cite{Kha24}. 
Despite this progress, the role of genuine quantum correlations in driven exclusion dynamics remains poorly understood, particularly in the nonequilibrium steady states (NESS) arising from the interplay between coherent and stochastic transport. 

In this work, we introduce a QCA realization of the TASEP and explore the competition of classical stochastic and coherent quantum transport.
Recent experiments in programmable Rydberg atom arrays have demonstrated scalable control of many-body quantum dynamics, placing the systems considered here within experimental reach~\cite{Labuhn2016, Bernien2017, Brow2020, Ebadi2021}.
We investigate the nonequilibrium phase diagram of this QCA-implementation of the quantum TASEP through large-scale simulations, and characterize the resulting NESS. 
Strikingly, although standard entanglement witnesses do not detect quantum correlations, the steady states exhibit clear quantum signatures beyond entanglement.
Given the simplicity and paradigmatic nature of the TASEP, our results indicate that quantum features beyond entanglement can appear even in minimal models of dissipative transport.\\
\begin{figure*}[t]
    \centering
    \includegraphics[width=\textwidth]{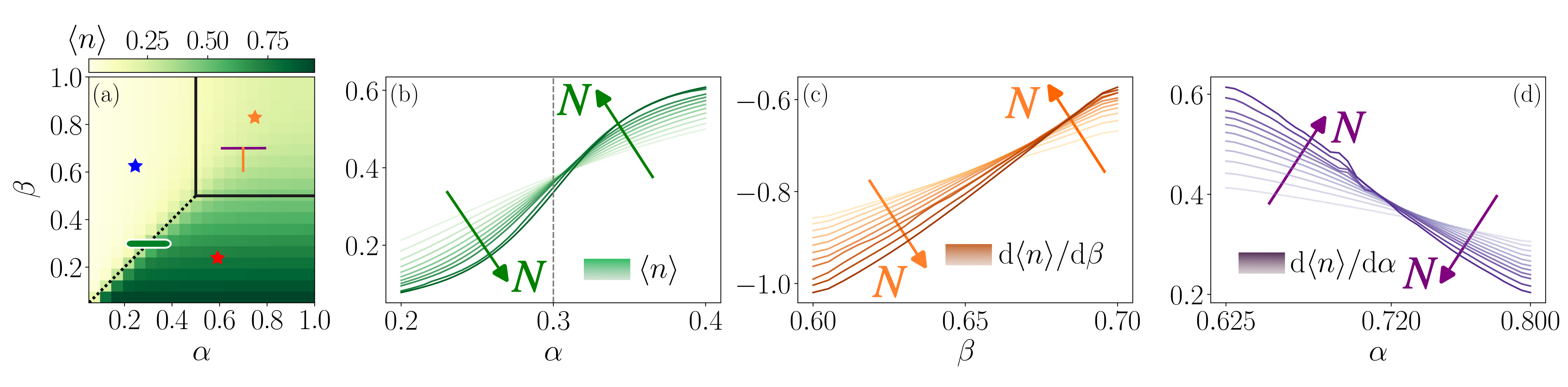}
    \caption{{\bf Average lattice occupation in finite-size NESS.} (a) Phase diagram for a system of $N=30$ lattice sites with coherent transport induced by the gate $U(\pi/4)$ and bulk hopping probability $\tau =0.75$. 
    The dashed line indicates the classical coexistence line separating the LD and HD phases and the solid line indicates the position of the transition to the MC phase in the classical model ($\omega =0$).
    Colored markers indicate parameter values we further study in Fig.~\ref{fig:LQU_and_Neg} and colored lines indicate parameter ranges for the order parameter scalings in panels (b)-(d). 
    Panel (b) shows the finite-size scaling of the average lattice occupation across the coexistence line ($\beta = 0.3$) and panels (c) and (d) show the finite-size scaling of the derivative of the average lattice occupation across the LD-MC ($\beta = 0.7$) and the HD-MC transition ($\alpha = 0.7$) respectively.
    All panels depict lattice sizes $N = 6,8,10,12,14,16,18,20,22,26,30$ encoded by the opacity of the lines with larger $N$ being more opaque. 
    }
    \label{fig:PDq}
\end{figure*}

{\bf Transport QCA model.---}We consider the quantum circuit model depicted in Fig.~\ref{fig:sketch}(a).
Each lattice site hosts a qubit, whose $Z$ eigenstates are interpreted as occupied $\ket{\bullet}$ or vacant $\ket{\circ}$, representing agents in a one-dimensional lattice.
Two neighboring qubits are paired with one ancillary qubit to form a {\it cluster}.
The automaton rule, realized by the quantum gate $G$, is applied on these clusters.
We explicitly consider a gate that consists of the subsequent application of a two-qubit gate $U$ followed by a three-qubit gate $D$, see Fig.~\ref{fig:sketch}(b).
The gate $U$ acts solely on the system qubits, whereas $D$ allows the ancilla to interact with the system qubits.
As depicted in Fig.~\ref{fig:sketch}(a), all ancilla qubits are measured and reinitialized once the automaton rule has been applied throughout the lattice, thereby inducing dissipative processes in the system.
For the clusters we introduce the eight quantum states $\kett{\threestate{\halfcirc}{\halfcirc}{\halfcirc}}$ where the ancilla label is chosen to sit on top of the two system labels, respecting the triangular structure introduced in Fig.~\ref{fig:sketch}.
Upon measuring the ancilla, the QCA rule implements a dissipative quantum channel in each system cluster which is described by two Kraus operators $K_{\halfcirc} = \brat{\threestate{\halfcirc}{\ }{\ }} D \kett{\threestate{\circ}{\ }{\ }}$~\cite{SM} corresponding to the measurement outcome.
This setup allows us to include a single dissipative processes per cluster in the otherwise coherent QCA.
An ancilla flip indicates if the corresponding dissipative process has occurred in a measured realization.
This scheme, can also be thought of as a repeated interaction framework~\cite{Karevski09}, in which successive layers of ancillary qubits interact with a system through the repeated application of the QCA rule, implemented by a quantum gate~\cite{Gill20,Gill21,Gill22,Gill23,Boneberg_2023}.

To explore quantum transport phenomena in this QCA framework, we choose the automaton rule $G$ as follows.
The two-qubit system gate is the parameterized gate~\footnote{As $U$ does not act on the ancilla, we discard its label here.}
\begin{align}
\begin{split}
U(\omega)
&=\ket{\circ\circ}\bra{\circ\circ}+ (\cos\omega \ket{\circ\bullet} - i\sin\omega\ket{\bullet\circ})\bra{\circ\bullet}\\
&+(\cos\omega \ket{\bullet\circ} - i\sin\omega \ket{\circ\bullet})\bra{\bullet\circ} + \ket{\bullet\bullet}\bra{\bullet\bullet}
\end{split}
\end{align}
which implements coherent nearest-neighbor hopping of excitations.
The system-ancilla component is 
\begin{align}
\begin{split}
D(\tau)
=\mathds{1} &+ i\sqrt{\tau}(\kett{\threestate{\circ}{\bullet}{\circ}}\brat{\threestate{\bullet}{\circ}{\bullet}} 
+\kett{\threestate{\bullet}{\circ}{\bullet}}\brat{\threestate{\circ}{\bullet}{\circ}})\\ 
&- (1-\sqrt{1-\tau})(\kett{\threestate{\bullet}{\circ}{\bullet}}\brat{\threestate{\bullet}{\circ}{\bullet}}+ \kett{\threestate{\circ}{\bullet}{\circ}}\brat{\threestate{\circ}{\bullet}{\circ}})
\end{split}
\end{align}
which implements the stochastic TASEP update rule upon tracing over the ancilla~\cite{SM}.
Since the ancilla is in the vacant state at the start of each time update, we see that, apart from the identity, only two terms in $D(\tau)$ contribute.
These correspond to a hopping transition term if the ancilla flips with probability $\tau$ ($\kett{\threestate{\bullet}{\circ}{\bullet}}\brat{\threestate{\circ}{\bullet}{\circ}}$) and a no-hop term if the ancilla does not flip
($\kett{\threestate{\circ}{\bullet}{\circ}}\brat{\threestate{\circ}{\bullet}{\circ}}$).
Hence, in the absence of the gate $U(\omega)$, the repeated ancilla interaction implements the bulk rules of a discrete-time TASEP with the bulk hopping probability $\tau$.
To realize open-boundary TASEP dynamics, we consider a boundary driven system~\cite{Prosen11,Karevski13,Caro18,Prosen11,Zni14,Zni14b,Ilievski2014}.
At the left boundary, excitations are injected with probability $\alpha$, while at the right boundary they are ejected with probability $\beta$.
These boundary processes can be implemented using two additional ancillary qubits that interact locally with the left and right boundary qubit respectively, in direct analogy to the ancilla-mediated bulk hopping, see the Supplemental Material for further details~\cite{SM}.

The stationary state for the classical TASEP can be analytically derived using a matrix product ansatz~\cite{Brank00,SM} and supports three distinct non-equilibrium phases.
A high density phase (HD) is separated from a low density phase (LD) by the coexistence line $\alpha = \beta$ up to $\alpha^\star = \beta^\star=1-\sqrt{1-\tau}$.
For $\alpha, \beta > \alpha^\star, \beta^\star$ the TASEP supports  a maximum current phase (MC).
While both the LD-MC and the HD-MC transition are continuous, the LD-HD transition is discontinuous~\cite{Brank00,SM}.
For the remainder of this work we choose $\tau = 0.75$.\\

{\bf Quantum NESS phases.---}We explore the influence of coherent transport induced by $U(\omega)$ on the phase diagram of the classical TASEP. 
We note that $U(\omega)$ is the identity in the zero- and the two-particle sector of each cluster and acts non-trivially only in the single-particle sector.
For $\omega = \pi/4$ this gate can create maximally entangled Bell states between neighboring system sites, such that the influence of the coherent transport is expected to be most significant.
Thus we restrict our analysis to $\omega=\pi/4$ for the remainder of this work.

We simulate the QCA model using a time-evolving block decimation (TEBD) tensor-network algorithm~\cite{Vidal04,Vidal07}, allowing the bond dimension to adapt dynamically during the evolution.
A single time step consists of the iterative application of the left boundary drive, the sequential bulk update,
and the right boundary drive, see Supplemental Material for additional details~\cite{SM}.
This procedure allows us to time-evolve the system to its NESS determined by the boundary driving parameters $\alpha$ and $\beta$.
Fig.~\ref{fig:PDq} shows our findings for systems up to $N=30$ qubits.

We see from Fig.~\ref{fig:PDq}(a) that the general shape of the classical phase diagram stays intact upon the introduction of strong coherent transport.
In Fig.~\ref{fig:PDq}(b)-(d) we illustrate the relevant order parameter crossings of the average lattice density that define the exact location of the phase boundaries.
Upon increasing the system size $N$ across the LD-HD coexistence line, the  average occupation decreases for smaller $\alpha$ and increases for larger $\alpha$. 
This behavior is indicated by a crossing of the finite-size curves and suggests the emergence of a jump discontinuity, see Fig.~\ref{fig:PDq}(b).
Conversely, the LD-MC as well as the HD-MC transition remain continuous and only the derivative of the average number density develops a discontinuity.
However, the extent of the MC phase is diminished compared to the classical case as can be seen from the finite-size scaling analysis in Figs.~\ref{fig:PDq}(c) and~(d).
As we explained above, the onset of the MC phase in the classical case depends on the numerical value of the bulk hopping probability $\tau$.
Larger $\tau$ values diminish the size of the MC phase, similar to what we observe in the presence of coherent transport.
Thus, the coherent transport appears to renormalize the bulk hopping probability towards higher values, which we further analyse in the Supplemental Material~\cite{SM}.
The collective behavior of the quantum model seems therefore largely governed by the underlying classical transport. 
Remarkably, we will show below that the NESS still exhibits clear quantum correlations.\\

\begin{figure*}[t]
\centering
    \includegraphics[width= .9\textwidth]{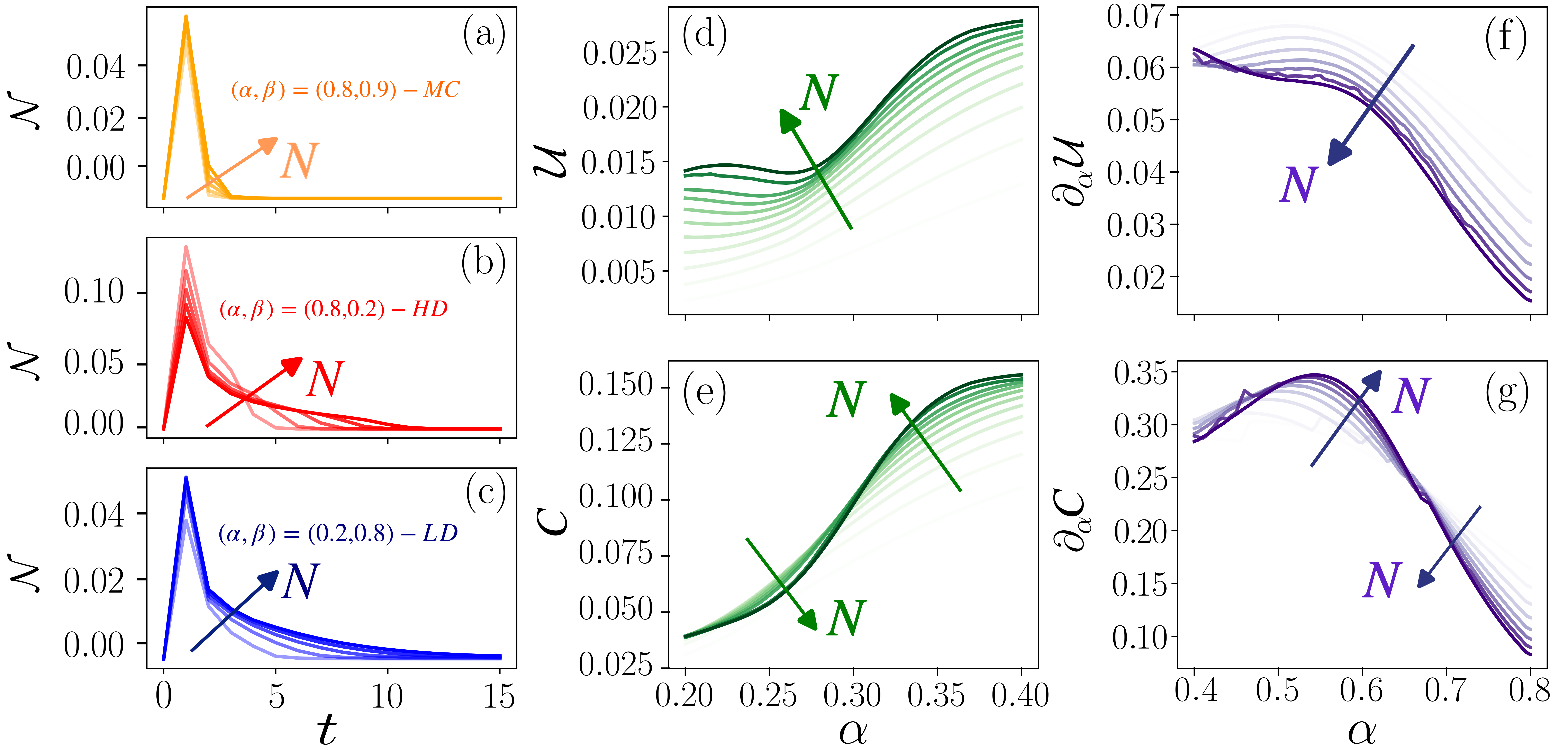}
    \caption{\textbf{Quantum correlations}.   (a)-(c) Time evolution of the half-system entanglement negativity for $N=6,8,10,12,14$ in the MC, HD and LD phase respectively (see markers in Fig.~\ref{fig:PDq}(a)).
    Panels (d) and (e) show the maximum two-site LQU and coherence with respect to the central bulk qubit across the HD-LD transition ($\beta=0.3$).
    Panels (f) and (g) show the derivative of the same quantities across the LD-MC transitions ($\beta =0.7$). All figures from panels (d)-(g) are plotted for $N=6,8,10,12,14,16,18,20,22,26,30.$
    }
    \label{fig:LQU_and_Neg}
\end{figure*}

{\bf Quantum correlations.---}We consider the build-up of quantum correlations during the time evolution.
The half-system negativity is defined as ${\cal N}= (||\rho^\Gamma||_1-1)/2 $, where $\rho^\Gamma$ denotes the partially transposed density matrix.
${\cal N}$ can witness entanglement via detection of violations of the positivity of the partial transpose $\rho^\Gamma$ (PPT criterion)~\cite{Vidal02,Plenio05}.
From Fig.~\ref{fig:LQU_and_Neg}(a)-(c) we generally observe that the negativity initially increases sharply in all regions of the phase diagram.
This is expected as initially quantum correlations are generated through coherent transport.
The behavior after the initial peak depends on the region of the phase diagram.
In the MC phase, the negativity vanishes rapidly following this initial spike and increasing system sizes do not support a longer-lived negativity.
Conversely in the LD and HD phase, as the system size increases, the initial negativity spike broadens, indicating that the corresponding entanglement survives for longer times when the system size is increased.
However, in the finite-size settings, the negativity eventually dies out and does not play a role in the correlation structure of the finite-size NESS.
We have also verified that all two-site reduced density matrix in the NESS are separable and that moment ratios of the partially transposed density matrix~\cite{Cara24} of the NESS for larger system sizes do not violate the PPT criterion~\cite{SM}, further substantiating the claim that generally the finite-size NESS correlations are not dominated by bipartite entanglement.
Our findings in the HD and LD phases indicate that a finite entanglement negativity may persist in the thermodynamic limit, because the time over which it survives increases with system size.
This contrasts with prior observations of weak correlation signatures in strongly driven systems~\cite{Temme_2012}, and reveals that substantial quantum correlations may persist in the steady state.

Since we find that the negativity consistently vanishes in the long-time limit in our simulations, it raises the question whether other types of quantum correlations may be present in the NESS.
Thus, we consider the local quantum uncertainty (LQU) ${\cal U}$~\cite{Giro13} which is closely related to quantum discord (see~\cite{Bera18} for a review) and a coherence measure $C$ in the classical basis that quantifies superpositions.
To evaluate these quantities, we first pick a central bulk site at site $c=N/2$ and compute all two-site reduced density matrices $\rho_{j,c}$. 
We then evaluate ${\cal U}$ by performing local measurements on the central qubit and $C$ by summing the absolute values of all off-diagonal elements of  $\rho_{j,c}$ and consider their maximum across the phase boundaries.
These measures are particularly appealing because they are experimentally accessible via two-site quantum state tomography~\cite{QSE,Auc11,Kim23,Ram21}.

The results are shown in Fig.~\ref{fig:LQU_and_Neg}(d)-(g).
Across the LD-HD transition we see that interestingly both $\mathcal{U}$ and $C$ clearly witness the transition by an uptake in absolute value around the transition point, see Figs.~\ref{fig:LQU_and_Neg}(d) and~(e).
However, both phases do show extensive $\mathcal{U}$ for the system sizes we investigated whereas $C$ indicates a crossing in the vicinity of the coexistence line.
Across the continuous LD-MC transition, $\partial_\alpha C$ shows a crossing in the vicinity of the critical point whereas the derivative $\partial_{\alpha}\mathcal{U}$ shows a strong drop upon entering the MC phase.
We observed similar results across the remaining HD-MC transition.

These results indicate that local two-site superpositions in the bulk are most pronounced in the MC phase and more pronounced in the HD phase than the LD phase.
This can be understood from the structure of the mixed state, whose dominant weight comes from pure states with nearest-neighbor empty–full pairings ($\kett{\circ\bullet},\kett{\bullet\circ}$), implying that local clusters are typically in the single-particle sector.
These states are mapped by the coherent transport $U$ into local superpositions whereas the zero- and two-particle sectors are not mapped into superpositions.
Therefore, the HD phase which has larger overlap with the single-particle sector than the LD phase also shows larger superpositions.
This explains the steep signal of the coherence and the LQU  across the LD-HD transition since the typical two-site configurations change drastically across the coexistence line.
Moreover, the MC phase is a homogeneous phase with the largest contribution from the single-particle sector and thus the most superpositions.
This indicates that our quantum witnesses do flatten out in the MC phase, yielding a diminished derivative as can be seen in Fig.~\ref{fig:LQU_and_Neg}.

\begin{figure}[t]
\centering
    \includegraphics[width= \columnwidth]{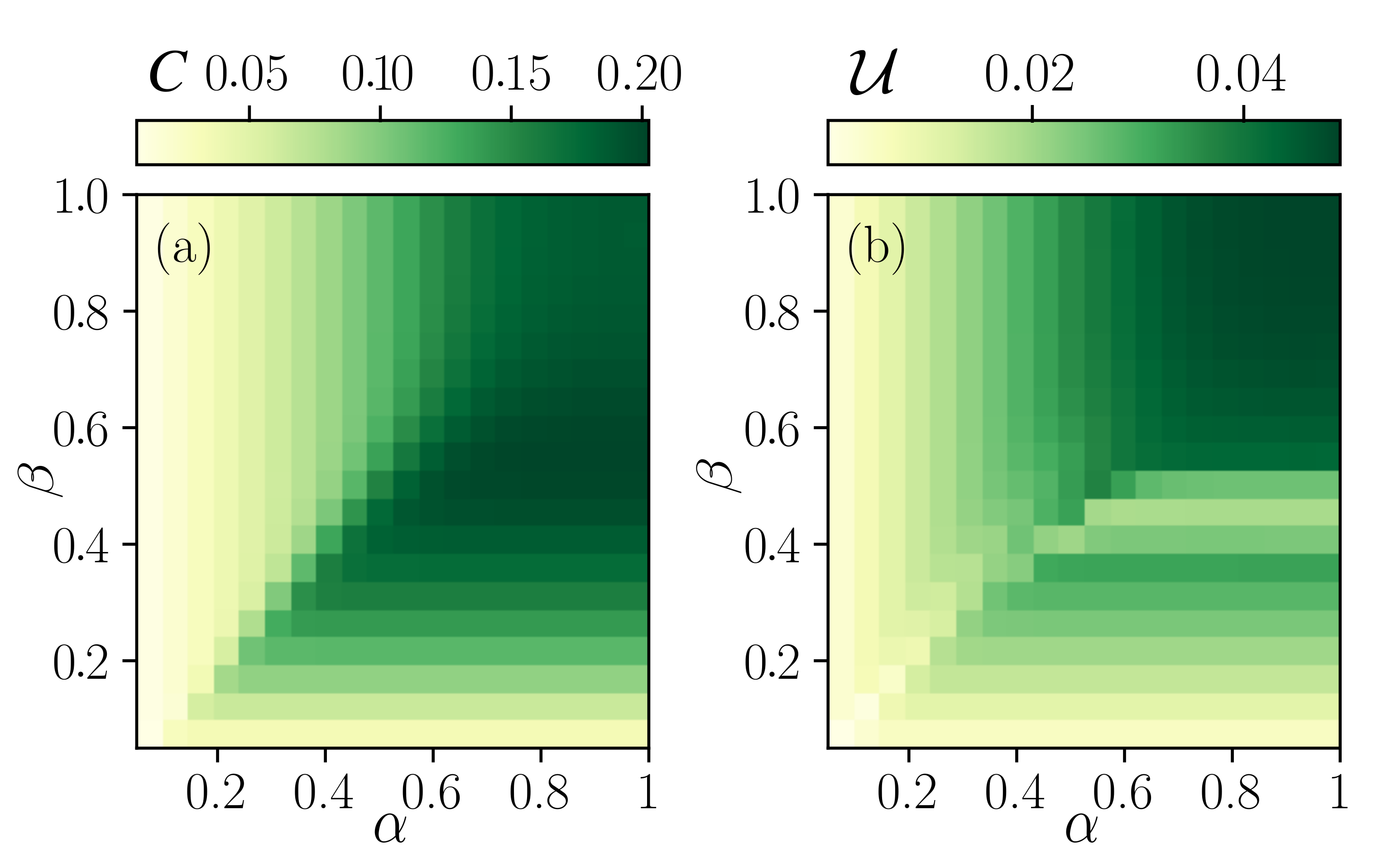}
    \caption{ {\bf Phase diagram from quantum correlations.} Panels (a) and (b) show  the two-site the coherence and LQU respectively, as described in the text, for the NESS that produced Fig.~\ref{fig:PDq}(a).}
    \label{fig:coh_lqu}
\end{figure}

Thus, the LQU and the coherence qualitatively reproduce the phase diagram from Fig.~\ref{fig:PDq}.
In Fig.~\ref{fig:coh_lqu} we show the results for ${\cal U}$ and $C$ for $N=30$ obtained from the matrix product states that also produced the occupation phase diagram in Fig.~\ref{fig:PDq}(a).
We see that both witnesses reproduce the structure of the quantum phase diagram, indicating that quantum correlations beyond entanglement are an essential feature of the NESS and are capable of detecting non-equilibrium phase transitions.\\

{\bf Conclusion.---}We introduced a QCA model for coherent and stochastic transport and analyzed a quantum extension of the paradigmatic TASEP.
While the quantum dynamics largely preserves the classical transport behavior at the level of local observables, leaving the phase diagram essentially unchanged, its correlation structure is fundamentally different.
We established that entanglement plays a crucial role in the approach to the NESS, with its decay sharply distinguishing between the different phases.
Strikingly, although no bipartite entanglement is detected in the NESS, the steady state exhibits clear quantum correlations beyond entanglement, including discord and coherence.
It would be interesting to further explore how these correlations affect the broader thermodynamic properties of the NESS~\cite{Stra17} and their impact on transport in realistic complex networks.
Our results show that nonclassical correlations persist in dissipative transport systems and even encode collective properties such as the phase structure in the absence of entanglement.
This establishes dissipative quantum transport as a setting where all quantum correlations remain relevant.
More broadly, our results suggest that strongly driven quantum systems generically host rich quantum correlation structures.\\

{\bf Data availability.---} The data displayed in the figures is available on Zenodo~\cite{zenodo}.\\

{\bf Acknowledgment.---}
The authors would like to thank Chris Hooley and Colin Rylands for useful discussions.
This work used the ARCHER2 UK National Supercomputing Service (https://www.archer2.ac.uk) as well as Coventry University's EPYC high performance computer.
S.W. acknowledges funding from the Engineering and Physical Sciences Research Council (EPSRC) under award UKRI3751.

\bibliography{refs}
\bibliographystyle{apsrev4-1}

\clearpage

\setcounter{equation}{0}
\setcounter{figure}{0}
\setcounter{table}{0}
\makeatletter
\renewcommand{\theequation}{S\arabic{equation}}
\renewcommand{\thefigure}{S\arabic{figure}}
\makeatletter
\onecolumngrid
\newpage

\setcounter{page}{1}
\begin{center}
{\Large SUPPLEMENTAL MATERIAL}
\end{center}
\begin{center}
\vspace{0.8cm}
{\Large \papertitle }
\end{center}
 \begin{center}
 Uddhav Sen,$^{1}$
 Federico Carollo,$^{2,1}$ and
 Sascha Wald$^{1}$
 \end{center}
 \begin{center}
 $^1${\em Centre for Fluid and Complex Systems, Coventry University, Coventry, CV1 2TT, United Kingdom}\\
 $^2${\em Dipartimento di Fisica, Sapienza Università di Roma, Piazzale Aldo Moro 5, 00185 Rome, Italy}
 \end{center}

\section{Kraus map}
In this section we derive the Kraus map that governs the QCA update rule.
We focus first on a single three-qubit cluster and then build the global map from this elementary building block.
Since the Kraus map for a cluster is obtained by tracing out the ancilla, we omit the system gate $U$ here for clarity.
For completeness, we recall the form of the three-qubit gate $D$:
\begin{align}
D=\mathds{1} + i\sqrt{\tau}(\kett{\threestate{\circ}{\bullet}{\circ}}\brat{\threestate{\bullet}{\circ}{\bullet}} 
+\kett{\threestate{\bullet}{\circ}{\bullet}}\brat{\threestate{\circ}{\bullet}{\circ}}) 
- (1-\sqrt{1-\tau})(\kett{\threestate{\bullet}{\circ}{\bullet}}\brat{\threestate{\bullet}{\circ}{\bullet}}+ \kett{\threestate{\circ}{\bullet}{\circ}}\brat{\threestate{\circ}{\bullet}{\circ}}) \ .
\end{align}
The Kraus operators are simply obtained by considering that the repeated interaction implies that the ancilla is always initialized in the vacant state $\ket{\circ}$.
However, after the system-ancilla interaction, the ancilla may be found either in the vacant state $\ket{\circ}$ or the occupied state $\ket{\bullet}$.
These two outcomes give rise to two Kraus operators $K_\circ$ and $K_\bullet$, labeled by the post-interaction ancilla state.
Hence, we find the cluster Kraus operators induced by the QCA update, viz.
\begin{align}
    K_\circ &= \brat{\threestate{\circ}{\ }{\ }} D \kett{\threestate{\circ}{\ }{\ }} = 
    \mathds{1} - (1-\sqrt{1-\tau})  \kett{\bullet\circ}\brat{\bullet\circ},\\
    K_\bullet &= \brat{\threestate{\bullet}{\ }{\ }} D \kett{\threestate{\circ}{\ }{\ }} = 
     i\sqrt{\tau} \kett{\circ\bullet}\brat{\bullet\circ} \ .
\end{align}
It is easy to verify that these operators form a valid Kraus map $K_\circ^\dagger K_\circ +  K_\bullet^\dagger K_\bullet = \mathds{1}$.
The inclusion of the unitary gate $U$ simply dresses these Kraus operators with a $U$ from the right.
The global update of the system density matrix can thus be written as
\begin{align} \label{evol}
    \rho(t+1) = \mathcal{K}_{R} \circ \mathcal{K}_{N-1} \circ \cdots \mathcal{K}_1 \circ \mathcal{K}_L [\rho(t)],
\end{align}
with $\mathcal{K}_j[ \cdot  ] = K^{(j)}_\circ  \cdot  K^{(j)\dagger}_\circ +  K^{(j)}_\bullet \cdot K^{(j)\dagger}_\bullet$ and $j$ indicating the bond $(j,j+1)$.
$\mathcal{K}_{R/L}$ being the right/left most boundary drive Kraus maps respectively.
These can be derived in a similar fashion considering the boundary sites interacting with an additional ancilla through the following gates respectively.
\begin{align}
    D_L &=\mathds{1} +i \sqrt{\alpha}(\kett{\twostate{\bullet}{\bullet}} \brat{\twostate{\circ}{\circ}}+\kett{\twostate{\circ}{\circ}} \brat{\twostate{\bullet}{\bullet}})-(1-\sqrt{
    1-\alpha})(\kett{\twostate{\circ}{\circ}} \brat{\twostate{\circ}{\circ}}+\kett{\twostate{\bullet}{\bullet}} \brat{\twostate{\bullet}{\bullet}}),\\
    D_R &= \mathds{1}+i \sqrt{\beta} (\kett{\twostate{\bullet}{\circ}} \brat{\twostate{\circ}{\bullet}}+\kett{\twostate{\circ}{\bullet}} \brat{\twostate{\bullet}{\circ}}) -(1- \sqrt{1-\beta}) (\kett{\twostate{\bullet}{\circ}} \brat{\twostate{\bullet}{\circ}}+\kett{\twostate{\circ}{\bullet}} \brat{\twostate{\circ}{\bullet}})\ .
\end{align}
Here, similar to the three-qubit gate, the ancilla label is chosen on top of the system label.

\section{Classical TASEP}
In this section we first review the solution of the steady state of the classical TASEP through a matrix product ansatz.
We focus on the discrete-time case with forward sequential update, since this scenario is considered in the main text. 
However, the ansatz can be easily adjusted for other update rules or to continuous time.
We then discuss the phase diagram of the TASEP through finite-size scaling obtained from the exact solution.

\subsection{Solution for the forward sequential update} 

\begin{figure*}[t]
\centering
    \includegraphics[width= 0.5\textwidth]{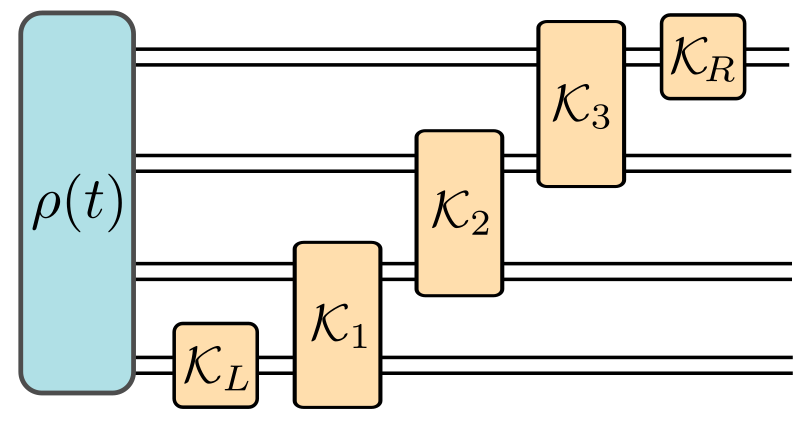}
    \caption{ {\bf Circuit Diagram for the sequential update rule.} The figure shows a circuit diagram for the sequential TASEP update with matrix product ansatz $\rho =\langle W| M^{\otimes N}|V\rangle$.}
    \label{fig:S1}
\end{figure*}

Here we sketch the derivation of the exact NESS $\rho_{\rm ss}$ of the classical TASEP with the forward sequential update as shown in Fig.~\ref{fig:S1}.
To this end, we make the ansatz $\rho_{\rm ss}= \bra{W} M \cdots M\ket{V}$ for a chain with $N$ sites.
Here $M$ is the repeated on-site tensor and $\bra{W}$ and $\ket{V}$ are boundary vectors.
A single time step is then given by the application of $\mathcal{K} = \mathcal{K}_R \circ \mathcal{K}_{N-1} \circ \cdots \circ \mathcal{K}_1 \circ \mathcal{K}_L$, see Fig.~\ref{fig:S1}.
If the NESS satisfies the following conditions
\begin{align} \label{DT:telescope}
    \bra{W}\mathcal{K}_L M = \bra{W}\hat{M}, \qquad
    \mathcal{K}_j \hat{M}M = M\hat{M},   \qquad
    \mathcal{K}_R \hat{M}\ket{V} = M\ket{V}
\end{align}
then the state is stationary by construction, i.e. $\mathcal{K} \rho_{\rm ss} = \rho_{\rm ss}$.
Hence, we may use these conditions to determine a self-consistent representation of the NESS.
It is important to note that this approach also technically holds for the quantum case but the resulting equations are significantly more complicated.
However, an analytical solution for the boundary driven case in the absence of bulk drive, i.e. $\tau=0$, does exist~\cite{Prosen11}.

In the classical case ($U_j = \mathds{1})$ we may express the tensor $M$ in terms of the physical dimension as $M = \operatorname{diag}(F, E)$. 
Here, $F$ and $E$ are matrices that we will determine and the fact that the system is classical allows us to neglect coherences, i.e. consider $M$ to be diagonal.
From the bulk condition ($0<j<N$) we then find
\begin{align} \label{eq:ana_class_bulk_deri}
\begin{split}
    \mathcal{K}_j (\hat{M} M)
    &=\tau (\sigma^-\hat{M}\sigma^+)   (\sigma^+ M \sigma^-)
    +  \hat{M} M 
    -(1-\sqrt{1-\tau}) \left[ (\hat{M}n) (M  \bar{n}) 
    +(n  \hat{M}) (\bar{n} M) 
    \right]\\
    &\qquad+ (1-\sqrt{1-\tau})^2 (n\hat{M}n) (\bar{n} M \bar{n})\\
    &= M \hat{M}
\end{split}
\end{align}
From Eq.~\eqref{eq:ana_class_bulk_deri} we can now derive straightforwardly
\begin{align} \label{eq:ana_class_bulk}
     [\hat{F},F] = [\hat{E},E] = 0, \qquad
     (1-\tau) \hat{F} E  = F\hat{E}, \qquad
     \hat{E}F + \tau \hat{F} E = E \hat{F}
\end{align}
The boundary conditions can be treated similarly and yield 
\begin{align}
    \label{eq:ana_class_bc}
    \bra{W}(F+\alpha E)=\bra{W}\hat{F}, \quad
     \bra{W} E (1-\alpha) = \bra{W} \hat E, \quad
     (1-\beta)\hat F\ket{V} = F\ket{V},\quad
     (\hat E + \hat F \beta)\ket{V} = E\ket{V}. 
 \end{align}

Eqs.~(\ref{eq:ana_class_bulk}) and~(\ref{eq:ana_class_bc}) can be solved by mapping the operator algebra to that of the classical PASEP (partially asymmetric simple exclusion process)~\cite{Blythe07}.
To this end we make the ansatz $\hat{F} = F - \lambda\mathds{1}$, $\hat{E} = E + \lambda\mathds{1}$ such that the bulk commutator conditions are trivially satisfied.
The remaining two bulk equations now coincide and similar for the boundary equations.
Hence, we are left with the following conditions
 \begin{align}
     \lambda F + \lambda (1-\tau) E = - \tau FE, \qquad
     \bra{W} E = -\frac{\lambda}{\alpha} \bra{W}, \qquad
      -\frac{\beta}{\lambda(1-\beta)} F \ket{V} = \ket{V}
 \end{align}
 Introducing rescaled matrices and parameters as in 
 $F' = F/\sqrt{1-\tau}$, $E' = \sqrt{1-\tau} E$, $\alpha' = \alpha/\tau$, $\beta' = \beta/\tau \cdot (1-\tau)/ (1-\beta)$ 
and fixing $\lambda = -\tau/\sqrt{1-\tau}$ maps our algebra to the PASEP algebra, viz.,
 \begin{align}
    F' +  E' =  F'E', \qquad 
    \bra{W} E' = \frac{1}{\alpha'} \bra{W},\ \qquad
     F' \ket{V} = \frac{1}{\beta'} \ket{V} .
 \end{align}
We may now use the well-established solution to the PASEP algebra ~\cite{Blythe07}.
To this end we introduce the infinite dimensional matrix ${\cal M} = \mathds{1} + \sum_{n=1}^\infty \ket{n}\bra{n+1} + \sqrt{(\alpha' + \beta' - 1)/\alpha' \beta'} \ket{0}\bra{1}$ and express the solution as 
\begin{align}\label{discrete_DE}
F = (1-\tau)^{\frac{1}{2}} \left( {\cal M} + (\beta'^{-1} - 1) \ket{0}\bra{0}   \right), \quad
E = (1-\tau)^{-\frac{1}{2}} \left( {\cal M}^T + (\alpha'^{-1} -1) \ket{0}\bra{0}  \right), \quad
\ket{V} = \ket{W} = \ket{0}.
\end{align}

\subsection{Phase diagram}
The analytical solution for the classical NESS that we presented in the previous section is an infinite bond dimension matrix product state.
We can truncate the bond dimension and evaluate the local occupations in order to determine the phase diagram of the TASEP. 
Here, as in the main test, we focus on the bulk hopping probability $\tau = 0.75$.
\begin{figure*}[t!]
    \centering
    \includegraphics[width=\textwidth]{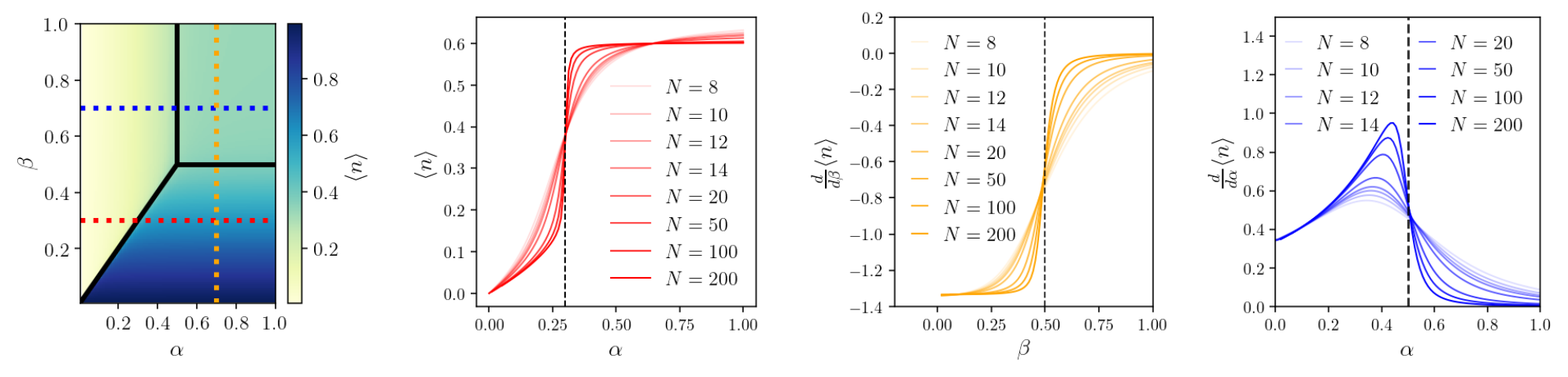}

    \caption{{\bf Classical discrete-time TASEP}.
    Panel (a) shows the NESS phase diagram of the classical TASEP with the sequential forward update in discrete-time. Data are obtained for a system of $N=200$ sites with a bond dimension $\chi = 20$ in the analytical NESS.
    Panel (b) shows the average chain filling for different system sizes across the red cut in panel (a) at $\beta = 0.3$.
    We see a clear crossing around $\alpha = \beta$ (vertical dashed line) indicating a discontinuous transition between LD and HD phase.
    Panels (c) and (d) show the derivative of the average chain filling across the orange ($\alpha = 0.7$) and blue ($\beta = 0.7$) cut respectively for different system sizes.
    We observe that the derivative in both cases shows a crossing indicating that both transitions from LD and HD to MC are continuous transitions.}
    \label{fig:classical}
\end{figure*}
In Fig.~\ref{fig:classical}(a) we show a heatmap plot of the average steady-state occupation.
We see that, the phase diagram is composed of three distinct regions:
\begin{itemize}
    \item a low-density region (LD) for $\alpha<\beta$ and $\alpha <1/2$,
    \item a high-density region (HD) for $\alpha>\beta$ and $\beta <1/2$,
    \item a maximum-current region (MC) for $1/2 < \alpha,\beta < 1$. 
\end{itemize}
The phase boundaries between these regions are well-established, see e.g. Ref.~\cite{Brank00}.

We proceed by carrying out a finite-size scaling analysis across the three phase boundaries.
Since the chosen update rule explicitly breaks particle-hole symmetry and it is established that correlations in the discrete-time systems may behave drastically different from their continuous-time counterparts~\cite{Wald2025} we wish to establish the nature of the transitions in the discrete-time TASEP.
Fig.~\ref{fig:classical}(b) shows the number density across the LD-HD cut for $\beta= 0.3$ and increasing system size $N$.
We clearly see an order parameter crossing and with increasing system size $N$ a jump discontinuity is developing at $\alpha_c = \beta^\star$.
Hence, in agreement with the continuous-time TASEP, the LD-HD transition is discontinuous and the $\alpha = \beta <1/2$ line is a coexistence line for the LD and HD phases.
Conversely, the HD-MC as well as the LD-MC transitions are continuous. 
In Fig.~\ref{fig:classical}(c) and~(d) we show exemplary cuts of the derivative (either with respect to $\alpha$ or $\beta$) $n'$ for $\alpha = 0.7$ and $\beta = 0.7$.
We see that, while the average occupation number is continuous across the MC phase boundaries, its derivative develops a jump discontinuity at the predicted critical region, where all lines intersect.

\section{The continuous-time limit} 

In this section we illustrate that the discrete-time dynamics we consider can also encompass the continuous-time case in the correct scaling limit.
Therefore, all probabilities of the stochastic processes need to scale with a time increment $dt$ which allows us to introduce the corresponding rate of the process.

In order to obtain the continuous-time limit of our dynamical map $\mathcal{K}$ we need to identify $\alpha = \gamma_L dt$, $\tau = \gamma dt$ and $\beta = \gamma_R dt$.
We then expand the Kraus map in powers of $dt$ up to linear order.
For example, for the left boundary this prescription yields
$
K_{\circ}^{(L)} \simeq \mathds{1}  - \frac{1}{2}\gamma_L dt (\mathds{1}-\ket{\bullet}\bra{\bullet}) +  {\cal O} (dt^2)$ and $
K_{\bullet}^{(L)} \simeq i\sqrt{\gamma_L dt}\ket{\bullet}\bra{\circ} + {\cal O} (dt)$.
Note that in order to obtain all linear terms we need to expand $K_{\circ}^{(L)}$ to linear order due to the mixed terms that the Kraus map generates but $K_{\bullet}^{(L)}$ only to $O(\sqrt{dt})$ as there are no cross terms.
We then obtain the usual Lindblad master equation for the pump at the left boundary, viz.,
\begin{align}
    \rho(t+dt) =  \rho(t) + \gamma_L \left( \sigma_0^+\rho(t) \sigma_0^- -\frac{1}{2} \{\sigma_0^- \sigma_0^+,\rho(t)\}\right) dt
\end{align}
with the ladder operators $\sigma^+ = \ket{\bullet}\bra{\circ}$ and $\sigma^- = \ket{\circ}\bra{\bullet}$.
The right boundary, as well as the bulk, can be treated in a similar fashion and we eventually find the full Lindbladian dynamics
\begin{align}
\begin{split}
    \frac{\partial\rho}{\partial t} = \mathcal{D}_{\rm TASEP}(\rho) \equiv \gamma_L \left( \sigma_0^+\rho(t) \sigma_0^- -\frac{1}{2}\{ \sigma_0^- \sigma_0^+,\rho(t)\} \right)
    + \gamma_R \left( \sigma_{N-1}^-\rho(t) \sigma_{N-1}^+ -\frac{1}{2} \{\sigma_{N-1}^+ \sigma_{N-1}^-,\rho(t)\}\right) \\
    +\sum_{j=1}^{N-1} \gamma \left( \sigma_{i-1}^-\sigma_{i}^+\rho(t)\sigma_{i-1}^+ \sigma_i^- -\frac{1}{2}\{ n_{i-1} (1-n_i),\rho(t)\} \right) \ .
\end{split}
\end{align}
Here, $n = \sigma^+ \sigma^-$ is the number operator.

\section{PPT criterion for the NESS for system size $N=30$}

\begin{figure}[b]
\centering
    \includegraphics[width= 0.36\textwidth]{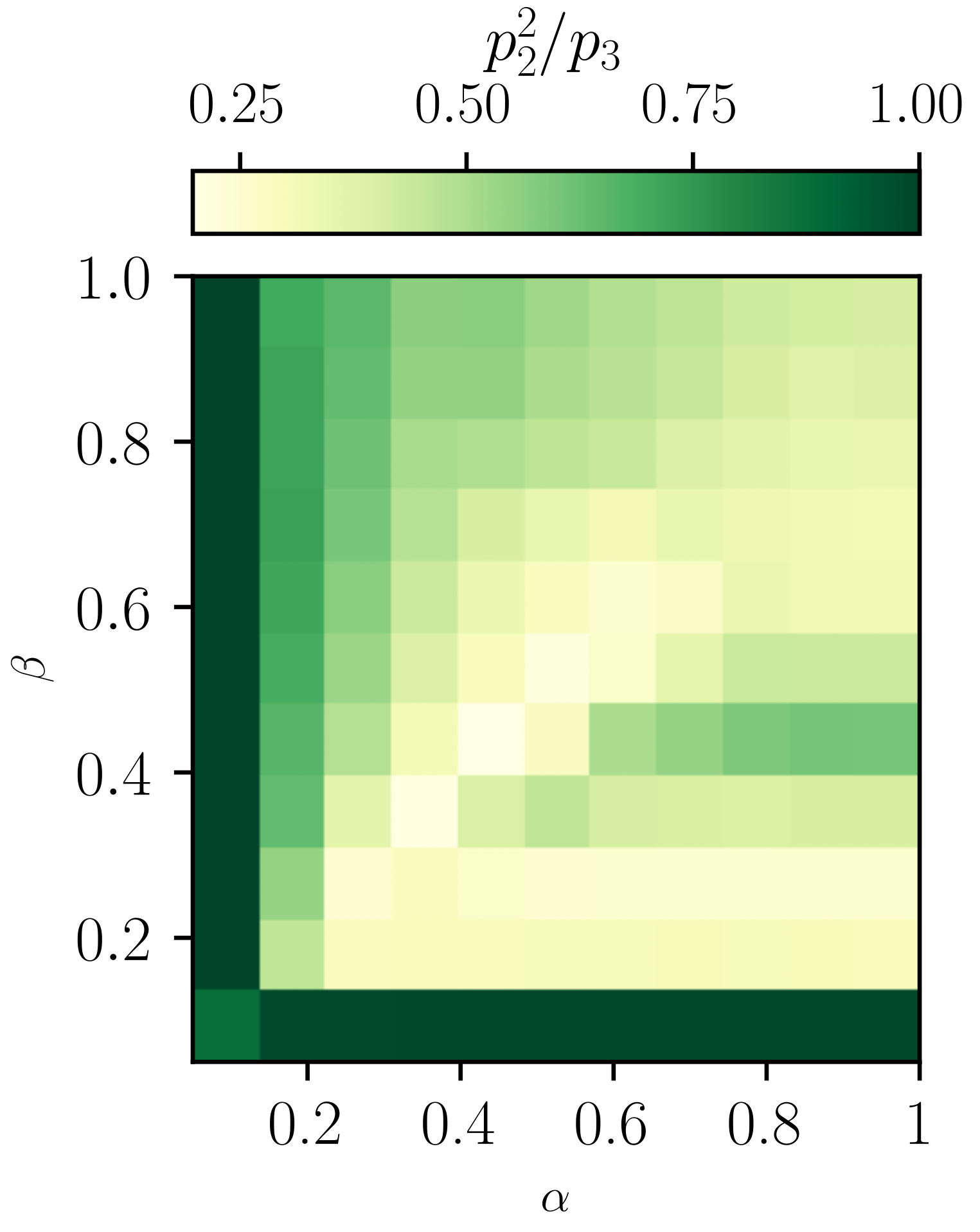}
    \caption{{\bf $p_2^2/p_3$ Phase Diagram for $N=30$}: The ratio  $p_2^2/p_3\le1$ everywhere implies no bipartite entanglement survives in the NESS. The ratio is exactly one at the edges as NESS is a pure state. }
    \label{fig:p2^2/p3}
\end{figure}

While it is challenging to evaluate entanglement witnesses such as the negativity for larger systems, it has been recently shown that moment ratios of the partial transpose can act as indicator for PPT violations.
More precisely Ref.~\cite{Elben20} has shown that $\rho$ is PPT implies that $1\geq p_2^2/p_3$ with $p_n = \operatorname{tr}((\rho^\Gamma)^n)$.
In this section we evaluate this moment ratio for the NESS with which we obtained the phase diagram in Fig.~\ref{fig:PDq}(a) for $N=30$.
The result is shown in Fig.~\ref{fig:p2^2/p3}.
We see that throughout the whole parameter range we consistently find $p_2^2/p_3 \leq 1$.
Hence, this criterion fails at detecting any PPT violations in the NESS and indicates that the NESS is at least valid with the PPT criterion~\cite{Horodecki09} which is consistent with our observation that bipartite entanglement seems to be absent in the NESS.
\section{Effective bulk hopping and phase boundaries for different coherence strengths $\omega$}
We mention in the main text that the presence of coherent transport seems to renormalize the bulk hopping probability towards larger effective values.
Here, we support this statement by explicitly evaluating the probability that the bulk gate $G$ moves a particle along the direction of the stochastic transport.
The gate $G$ can be written as
\begin{align}
\begin{split}
    G&= U + i \sqrt{\tau} \kett{\threestate{\circ}{\bullet}{\circ}} \left(\cos \omega \brat{\threestate{\bullet}{\circ}{\bullet}} - i \sin \omega \brat{\threestate{\bullet}{\bullet}{\circ}} \right) + i \sqrt{\tau} \kett{\threestate{\bullet}{\circ}{\bullet}}\left(\cos \omega \brat{\threestate{\circ}{\bullet}{\circ}} - i \sin \omega \brat{\threestate{\circ}{\circ}{\bullet}}\right)  \\
&-(1-\sqrt{1-\tau})\kett{\threestate{\bullet}{\circ}{\bullet}}\left(\cos \omega \brat{\threestate{\bullet}{\circ}{\bullet}} - i \sin \omega \brat{\threestate{\bullet}{\bullet}{\circ}}\right) -(1-\sqrt{1-\tau})\kett{\threestate{\circ}{\bullet}{\circ}}\left(\cos \omega \brat{\threestate{\circ}{\bullet}{\circ}} - i \sin \omega \brat{\threestate{\circ}{\circ}{\bullet}}\right)
\end{split}
\label{eq:SM_G}
\end{align}
First we consider the dissipative transition amplitude $\brat{\threestate{\bullet}{\circ}{\bullet}} G\kett{\threestate{\circ}{\bullet}{\circ}}=i\sqrt{\tau} \cos \omega$.
Here, the ancilla flip guarantees that the dissipative bulk hopping is taking place.
The corresponding transition probability is
\begin{align}
 \big|\brat{\threestate{\bullet}{\circ}{\bullet}}G \kett{\threestate{\circ}{\bullet}{\circ}}\big|^2
 &= \tau \cos^2 \omega.
\end{align}
However, the coherent transport also allows for transport in the direction of the stochastic transport without ancilla flip, i.e.,
\begin{align}
 \big|\brat{\threestate{\circ}{\circ}{\bullet}}G \kett{\threestate{\circ}{\bullet}{\circ}}\big|^2
 =\big|\brat{\circ \bullet} U \kett{\bullet \circ }\big|^2
 =\sin^2\omega.
\end{align}
Hence, the local cluster attains an effective forward hopping probability which is generically larger than in the absence of coherent transport:
\begin{align}
    p_{\bullet\circ\to\circ\bullet} = 
    \big|\brat{\threestate{\bullet}{\circ}{\bullet}}G \kett{\threestate{\circ}{\bullet}{\circ}}\big|^2
    +
     \big|\brat{\threestate{\circ}{\circ}{\bullet}}G \kett{\threestate{\circ}{\bullet}{\circ}}\big|^2
     =
     \sin^2\omega + \tau \cos^2 \omega
     = \tau + (1-\tau)\sin^2\omega >\tau .
\end{align}

\begin{figure}[t!]
    \centering
    \includegraphics[width=\textwidth]{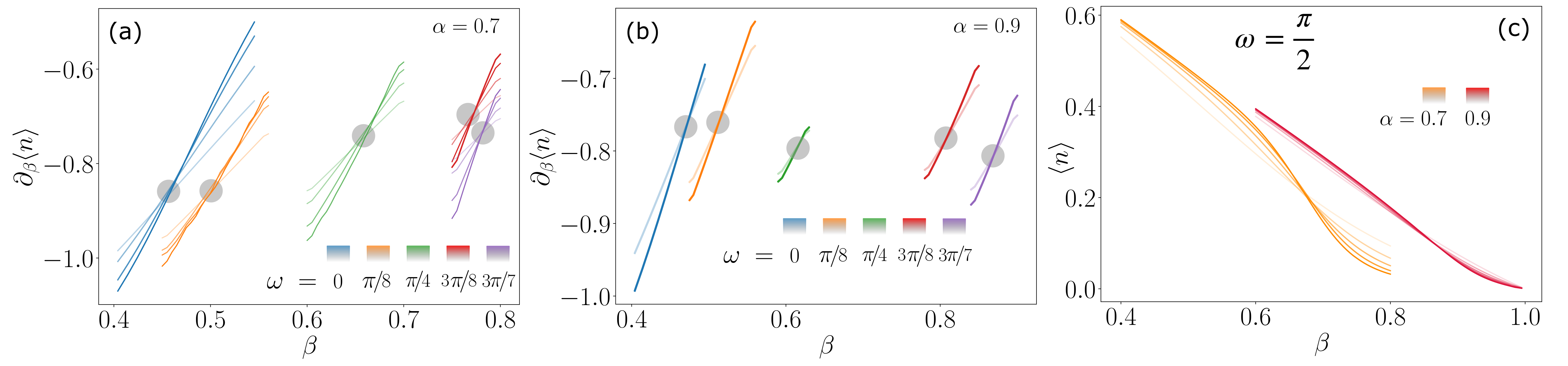}
    \caption{{\bf HD-MC phase boundary estimation for different strength of coherent transport.} Panel (a) shows derivative of the average chain filling across $\beta$ for increasing $\omega$ for  $\alpha=0.7$. This indicates that the MC phase is reduced but does not seem to disappear with increase in $\omega$. The gray circles highlight the crossing region. Data are obtained for a system of $N=6, 10, 16, 20$.
    Panel (b) shows derivative of the average chain filling across $\beta$ for increasing $\omega$ for  $\alpha=0.9$. This was done in order to check if results found in panel (a) is consistently seen throughout the HD-MC crossing and check if the MC phase disappears. The gray circles highlight the crossing region. Data are obtained for a system of $N= 10, 16$.
    Panel (c)  shows the average chain filling across the orange ($\alpha = 0.7$) and red ($\alpha = 0.9$) cut respectively for $\omega=\pi/2$. We look at system sizes $N=10,20,30,40,50$ and observe that the MC phase seems to have disappeared as both the cuts show a jump transition at around $\alpha=\beta$.}
    \label{fig:SM_phase_boundary}
\end{figure}

To further develop understanding of the phase boundaries for different coherence strengths $\omega$, we consider cuts from the high density to the maximum current phase for $\omega =0, \pi/8, \pi/4, 3\pi/8, 3\pi/7 $, see Fig.~\ref{fig:SM_phase_boundary}.

We clearly see that as $\omega$ increases, the value of $n$ that locates the phase boundary drifts upwards.
At $\omega = \pi/2$ we expect the MC phase to completely vanish since the effective bulk transport probability becomes $ p_{\bullet\circ\to\circ\bullet} = 1$ and it is well-established that deterministic bulk transport ($\tau = 1$) deteriorates the MC phase~\cite{Blythe07}.
This expectation is supported by the numerical observations in Fig.~\ref{fig:SM_phase_boundary} and can be substantiated by comparing the dynamics for $\tau = 1, \ \omega = 0$ and $\tau =0, \ \omega=\pi/2$.
Due to the forward update of our dynamics, both parameter combinations produce the same dynamics for $\beta = 1$ and an initially empty system.
However, while the former dynamics strictly only allows forward hopping, the latter does allow for backward hopping. 
Nevertheless, the forward update suppresses this backflow such that we expect the large-scale collective properties to not be influenced.

\end{document}